\documentclass{eas}
\usepackage{graphicx}

\begin{document}

\title{The Elemental Abundance Distributions of Milky Way Satellite
  Galaxies}\thanks{Data herein were obtained at the W.~M. Keck
  Observatory, which is operated as a scientific partnership among the
  California Institute of Technology, the University of California,
  and NASA.  The Observatory was made possible by the generous
  financial support of the W.~M. Keck Foundation.}

\author{Evan~N.~Kirby}
\address{Hubble Fellow, California Institute of Technology, Department
  of Astronomy, 1200 E.\ California Blvd., MC~249-17, Pasadena, CA
  91125, \email{enk@astro.caltech.edu}}

\runningtitle{Kirby: Abundance Distributions in MW dSphs}

\begin{abstract}
The chemical compositions of the stars in Milky Way (MW) satellite
galaxies reveals the history of gas flows and star formation (SF)
intensity.  This talk presented a Keck/DEIMOS spectroscopic survey of
the Fe, Mg, Si, Ca, and Ti abundances of nearly 3000 red giants in
eight MW dwarf satellites.  The metallicity and alpha-to-iron ratio
distributions obey the following trends: (1) The more luminous
galaxies are more metal-rich, indicating that they retained gas more
efficiently than the less luminous galaxies.  (2) The shapes of the
metallicity distributions of the more luminous galaxies require gas
infall during their SF lifetimes.  (3) At $\mathrm{[Fe/H]} < -1.5$,
[$\alpha$/Fe] falls monotonically with increasing [Fe/H] in all MW
satellites.  One interpretation of these trends is that the SF
timescale in any MW satellite is long enough that Type Ia supernovae
exploded for nearly the entire SF lifetime.
\end{abstract}

\maketitle

\section{Introduction}
\label{sec:intro}
How do dwarf galaxies form their stars?  How much gas was accreted by
gravitational attraction?  How much gas left the galaxy from supernova
winds (Dekel \& Silk \cite{dek86}) or tidal or ram pressure stripping
(Lin \& Faber \cite{lin83}) from interaction with the Milky Way (MW)?

The most basic approach to answering these questions is to fit an
analytic model of chemical evolution to the observed metallicity
distribution function (MDF).  The star formation histories (SFHs) may
also be deduced from the colors and magnitudes of the population and
from stellar spectroscopy.  Photometrically derived SFHs are most
sensitive to young, metal-rich stars because the separation between
isochrones increases with decreasing age and increasing metallicity.
Elemental abundances obtained from spectroscopy do not give absolute
ages, but they can provide finer relative time resolution for old,
metal-poor populations.  The trend of the alpha-to-iron ratio with
iron abundance, a proxy for elapsed time or integrated star formation
(SF), reveals the relative SFH with a resolution of about 10~Myr, the
approximate timescale for a Type~II SN.


With the aim of quantifying the SFHs of Local Group dwarf galaxies, my
collaborators and I have collected a large number of Keck/DEIMOS
spectra of individual red giants in MW dwarf satellite galaxies.  Our
catalog of abundances (Kirby \etal\ \cite{kir10}) is based on spectral
synthesis of the medium-resolution DEIMOS spectra.  The catalog
contains nearly 3000 red giants in eight MW dwarf spheroidal (dSph)
galaxies: Fornax, Leo~I and II, Sculptor, Sextans, Draco, Canes
Venatici~I, and Ursa Minor.  The number of stars in each dSph ranges
from 124 (Sextans) to 825 (Leo~I).  It is the largest homogeneous
chemical abundance data set in dwarf galaxies.  The biggest advantage
of our data set is that all of the spectra were obtained with the same
spectrograph configuration, and all of the abundances were measured
with the same spectral synthesis code.

\section{MW Dwarf Galaxy Metallicity Distributions}
\label{sec:mdfs}

\subsection{Luminosity-Metallicity Relation}

\begin{figure}
\centering
\resizebox{0.55\columnwidth}{!}{%
  \includegraphics{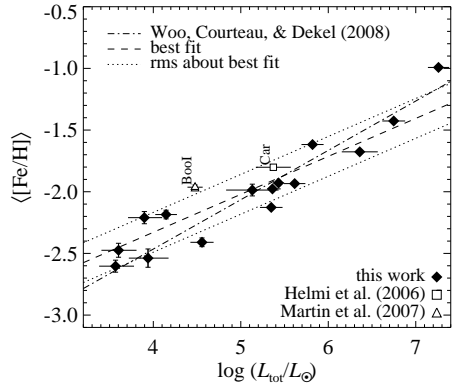} }
\caption{The mean [Fe/H] of MW dSphs as a function of total
  luminosity.  The diamonds represent our spectral synthesis
  measurements.  Open symbols are measurements based on the equivalent
  width of the Ca triplet.  Woo \etal's (\protect \cite{woo08})
  relation from Local Group galaxies includes galaxies much more
  luminous than Fornax.}
\label{fig:lzr}
\end{figure}

The average metallicities of more luminous dwarf galaxies are larger
than for fainter dwarf galaxies.  Figure~\ref{fig:lzr} shows the
luminosity-metallicity relation (LZR) for dwarf galaxies based on our
spectral synthesis measurements, along with Ca triplet-based
measurements for Carina and Bo{\" o}tes~I.  This LZR includes all MW
dwarfs less luminous than Sagittarius except the least luminous
objects (Willman~1; Segue~1, 2, and 3; Bo{\" o}tes~II; Leo~V; and
Pisces~I and II).  The following equation describes the orthogonal
regression fit:

\begin{equation}
\langle{\rm [Fe/H]}\rangle = (-2.02 \pm 0.04) + (0.31 \pm 0.04) \log \left(\frac{L_{\rm tot}}{10^5 L_{\odot}}\right) \: . \label{eq:lzrall}
\end{equation}

A straight line may be an overly simplistic model to the
luminosity-metallicity relation.  The dwarfs with $\log (L/L_\odot) >
5.2$ seem to lie along a steeper line than the less luminous dwarfs.
In order to better show that difference, the dot-dashed line in
Fig.~\ref{fig:lzr} is the LZR for more luminous Local Group galaxies
(Woo \etal\ \cite{woo08}).  It is an excellent fit to the luminous
half of those 17 dwarfs.  This is not surprising because many of those
dwarfs were included in Woo \etal's sample.  However, the fit is not
good to the dwarfs with $\log (L/L_\odot) < 5.2$.

\subsection{Analytic Chemical Evolution Models}
\label{sec:analyticmodels}

\begin{figure}
\centering
\begin{minipage}[t]{0.51\columnwidth}
\centering
\resizebox{\columnwidth}{!}{%
  \includegraphics{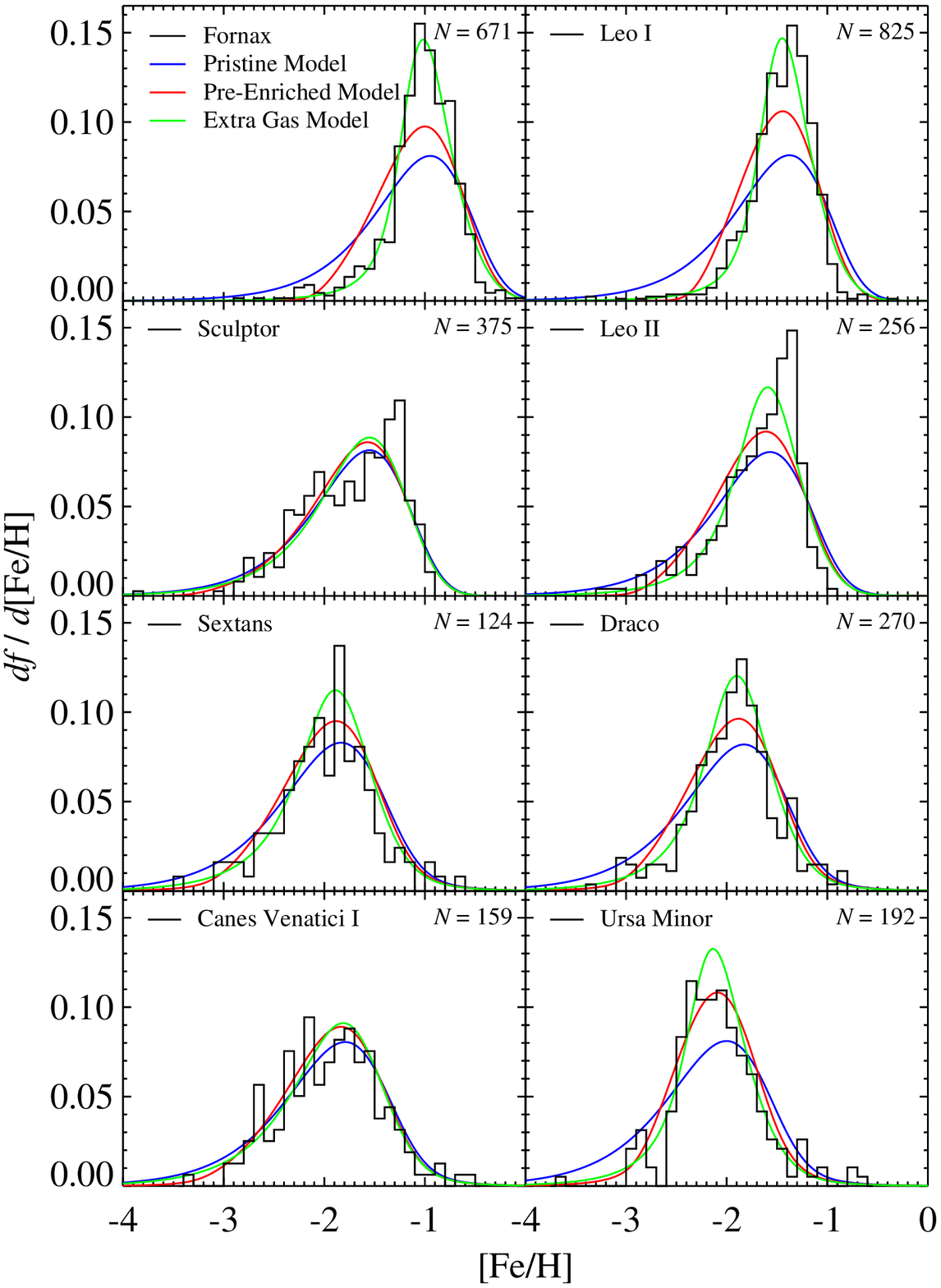} }
\caption{The metallicity distributions of dSphs expressed as a
  fraction of the total number of observed stars.  The panels are
  arranged from left to right and then top to bottom in decreasing
  order of dSph luminosity.  The blue, red, and green curves are the
  maximum likelihood fits to galactic chemical evolution models
  convolved with the measurement uncertainties.}
\label{fig:fehhists}
\end{minipage}
\hfil
\begin{minipage}[t]{0.46\columnwidth}
\centering
\resizebox{\columnwidth}{!}{\includegraphics{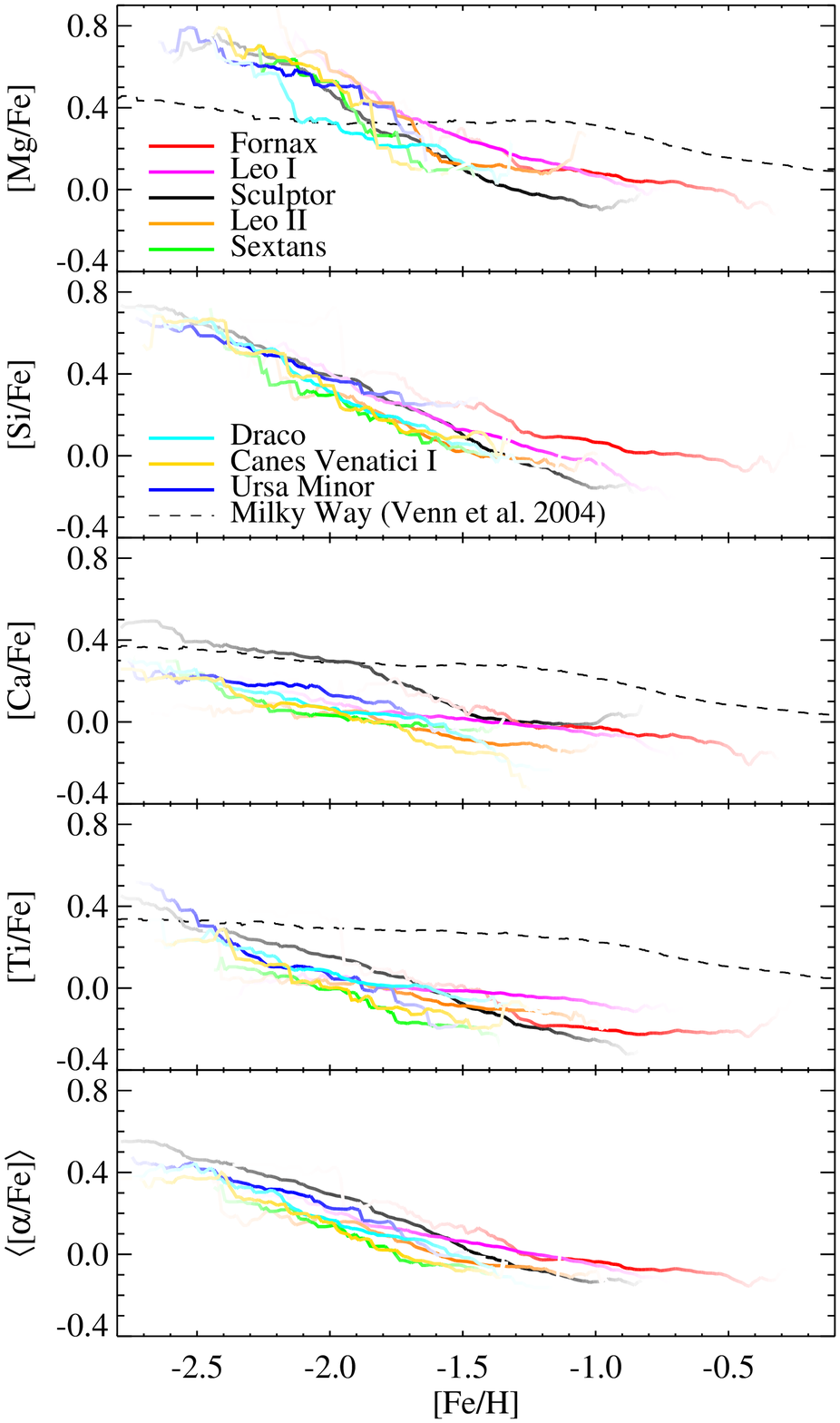}}
 \caption{The moving averages of abundance ratios for the eight dSphs
   and the MW (Venn \etal\ \cite{ven04}).  The bottom panel shows the
   average of the other panels.  The line weight is proportional to
   the number of stars contributing to the average.  The legend lists
   the dSphs in decreasing order of luminosity.}
\label{fig:alphatrends}
\end{minipage}
\end{figure}

Figure~\ref{fig:fehhists} shows the observed MDFs of each of the eight
dSphs, along with the best fits for three analytic chemical evolution
models.  A Leaky Box Model starting from zero-metallicity gas does not
faithfully describe any of the galaxies because it encounters the same
``G dwarf problem'' that once complicated the interpretation of the
MW's metallicity distribution.  A model with a fairly arbitrary
prescription for an increase in gas supply better describes the shape
of the MDFs by allowing for a narrower peak and a longer metal-poor
tail than the Leaky Box Model.  Permitting an initial metallicity
(pre-enrichment) allows the shape of the Leaky Box Model to better fit
the observed MDFs, but only for Canes Venatici~I does the Pre-Enriched
Model fit obviously better than the Extra Gas Model.  In several
cases, the Extra Gas Model fits much better than the Pre-Enriched
Model.

The dSphs may be separated into two broad categories: more luminous,
infall-dominated (Fornax, Leo~I, and Leo~II) and less luminous,
outflow-dominated (Sextans, Ursa Minor, Draco, and Canes Venatici~I).
The more luminous, infall-dominated dSphs are more consistent with the
Extra Gas Model than the Pristine Model or the Pre-Enriched Model.
The less luminous, outflow-dominated dSphs show similar low effective
yields ($0.007 \le p \le 0.018$) compared to the more luminous dSphs.
One explanation for the low values of $p$ is that gas outflow reduces
the effective yield below the value achieved by SN ejecta.  The low
masses of the less luminous dSphs rendered them unable to retain their
gas.  Gas flowed out of the galaxies from internal mechanisms, such as
SN winds, and external mechanisms, such as ram pressure stripping.
The outflows prevented the MDFs from achieving a high
$\langle\mathrm{[Fe/H]}\rangle$ and caused the MDFs to be more
symmetric than the more luminous dSphs.

We surmise that dSph luminosity is a good indicator of the ability to
retain or accrete gas.  For the more luminous dSphs, an increase in
the gas reservoir during the SF lifetime shapes the MDF and keeps the
effective yield and therefore the mean metallicity high.  The less
luminous dSphs are less able to retain gas that leaves via supernova
winds or interaction with the MW.  Finally, all of the analytic
chemical evolution models we consider are much too narrow to explain
the MDF for Sculptor.  However, a numerical model that treats iron as
a secondary nucleosynthetic product can reproduce Sculptor's MDF very
well (see Fig.~\ref{fig:scl}).

\section{MW Dwarf Galaxy [$\alpha$/Fe] Distributions}
\label{sec:afe}

\subsection{Universal Abundance Pattern in dSphs}

\begin{figure}[t!]
\centering \columnwidth=.45\columnwidth
\resizebox{\columnwidth}{!}{\includegraphics{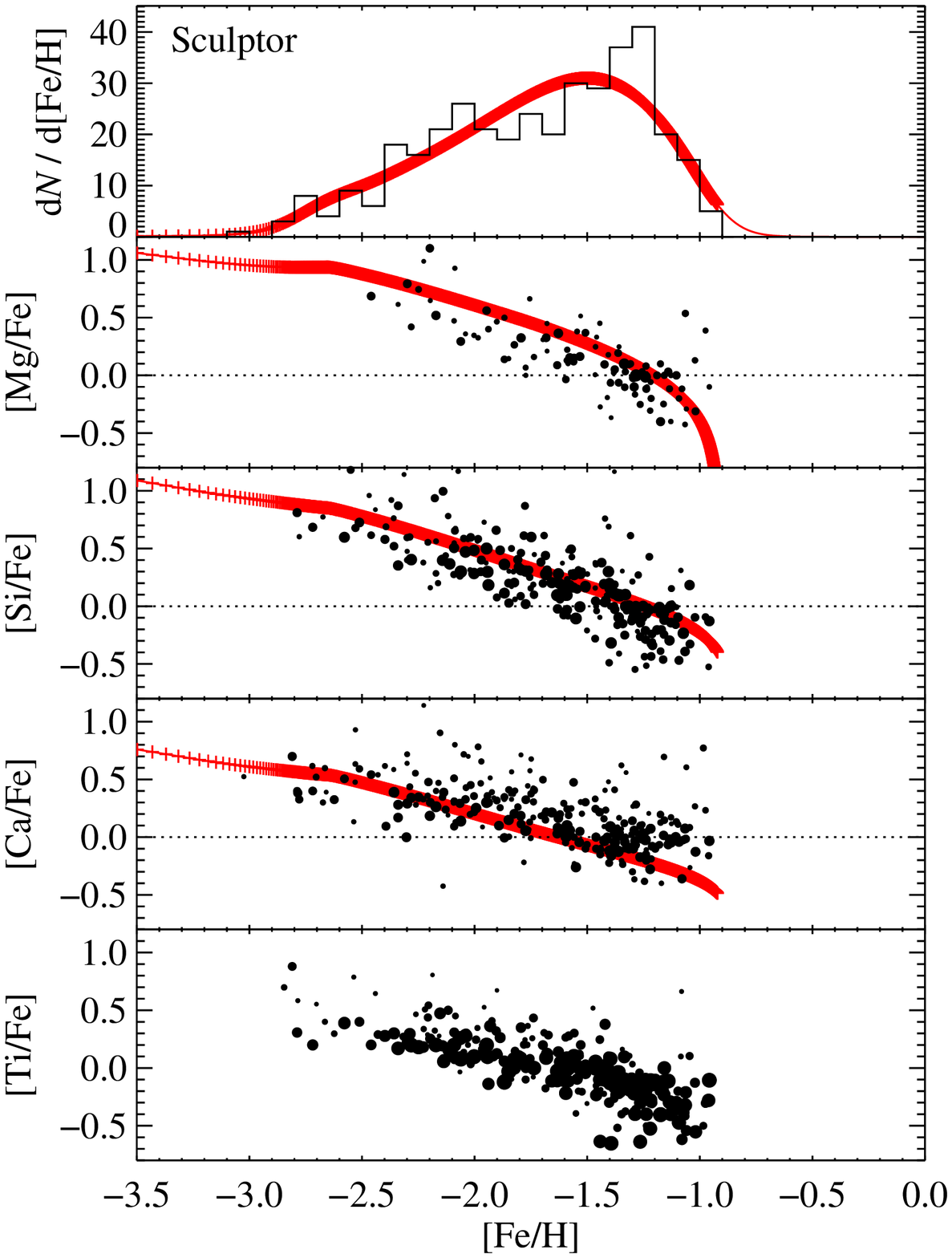}}
\hfil
\resizebox{\columnwidth}{!}{\includegraphics{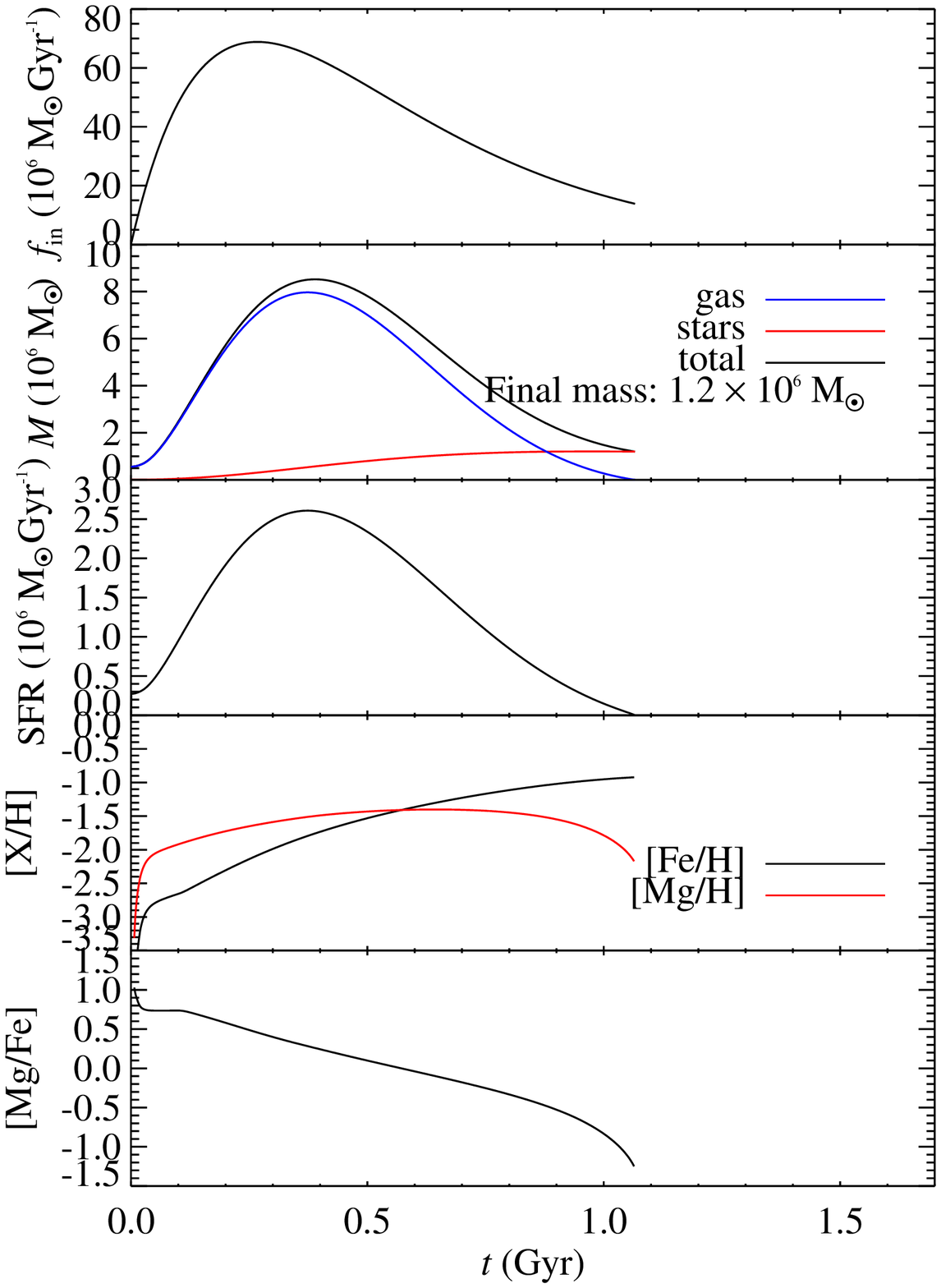}}
 \caption{The observed abundance ratios and the best-fit gas flow and
   SFH model for Sculptor.  {\it Left:} The observed MDF and
   [$\alpha$/Fe] distribution (\textit{black}) and the model
   (\textit{red}).  We do not show the model results for [Ti/Fe]
   because the theoretical SN yields are inaccurate.  {\it Right:} The
   gas flow and star formation history for the best-fit model.  From
   top to bottom, the panels show the gas inflow rate; the stellar,
   gas-phase, and total baryonic mass; the SF rate; the Fe and Mg
   abundances; and [Mg/Fe].}
\label{fig:scl}
\end{figure}

Figure~\ref{fig:alphatrends} shows the trend lines of the different
element ratios with [Fe/H].  The abundance distributions of dSphs
evolve remarkably similarly.  Although the dSphs span different ranges
of [Fe/H], $\langle[\alpha/\rm{Fe}]\rangle$ follows roughly the same
trend line.  This similarity contradicts the reasonable expectation
that different dSphs should show a knee in [$\alpha$/Fe] at different
values of [Fe/H] (Matteucci \& Brocato \cite{mat90}; Gilmore \& Wyse
\cite{gil91}; Tolstoy \etal\ \cite{tol09}).  In fact, Venn \& Hill
(\cite{ven08}) do indeed find a knee at $\mathrm{[Fe/H]} = -1.8$ in
their preliminary measurements for [Ca/Fe] in Sculptor.  Our
measurements of [Ca/Fe] in Sculptor also show a knee at the same
metallicity and the same [Ca/Fe].  The element ratios that would
better identify the onset of Type~Ia SNe, [Mg/Fe] and [Si/Fe], do not
show a knee for any dSph.

The lack of knees for $\mathrm{[Fe/H]} > -2.5$ and the lack of
low-metallicity plateaus in the [$\alpha$/Fe] distributions implies
that Type~Ia SNe exploded throughout almost all of the SFHs of all
dSphs.  Of course, the very first stars, which have yet to be found,
must be free of all SN ejecta.  The stars to form immediately after
the first SNe must incorporate only Type~II SN ejecta.  The very
lowest metallicity stars in dSphs likely represent this population.
Stars with $\mathrm{[Fe/H]} \ge -2.5$ formed after the Type~Ia
SN-induced depression of [$\alpha$/Fe].

\subsection{Numerical Chemical Evolution Model for Sculptor}
\label{sec:numericalmodels}

In order to provide a rough interpretation of the abundance trends in
Sculptor, we have developed a rudimentary model of chemical evolution.
The model tracks the SF rate, Types~II and Ia supernova explosions,
and supernova feedback.  Figure~\ref{fig:scl} shows the results of the
model compared to our observations.  We could not reproduce the width
of Sculptor's MDF with an analytic model of chemical evolution.  Our
more sophisticated model, which more properly treats Fe as a delayed
nucleosynthetic product with multiple origins (Types~II and Ia SNe),
yields a broad, well-matched MDF for the appropriate choice of
parameters.

Majewski \etal\ (\cite{maj99}) found that Sculptor undoubtedly
contains multiple stellar populations based on its horizontal and red
giant branch morphologies.  The existence of a metallicity spread, the
depression of [$\alpha$/Fe] with increasing metallicity, and the
radial change in horizontal branch morphology (Tolstoy
\etal\ \cite{tol03}; Babusiaux \etal\ \cite{bab05}) means that SF
lasted for at least as long as the lifetime of a Type~Ia SN and likely
for a few Gyr.  Our chemical evolution model, with a SF duration of
1.1~Gyr, conforms to this description of Sculptor's SFH.

The central regions of Sculptor are dominated by a more metal-rich
population than the outer regions (Battaglia \etal\ \cite{bat08}).
Our sample is centrally concentrated in order to maximize the sample
size.  The selection results in a bias toward metal-rich, presumably
younger stars, possibly shortening the derived the SF duration
compared to what we would deduce from a more radially extended sample.

Support for this work was provided by NASA through Hubble Fellowship
grant 51256.01 awarded by the Space Telescope Science Institute, which
is operated by the Association of Universities for Research in
Astronomy, Inc., for NASA, under contract NAS 5-26555.

\end{document}